\documentclass[preprint,aps,eqsecnum]{revtex4-1}

\usepackage{amsmath,amssymb,amsfonts,dcolumn,color,graphicx,graphics,
latexsym,placeins,epsfig}
\usepackage{subfigure,rotating,hyperref}

\newcommand{\be}{\begin{equation}}
\newcommand{\ee}{\end{equation}}
\newcommand{\ba}{\begin{eqnarray}}
\newcommand{\ea}{\end{eqnarray}}

\usepackage[normalem]{ulem}
\newcommand{\stkout}[1]{\ifmmode\text{\sout{\ensuremath{#1}}}\else\sout{#1}\fi}

\begin{document}

\title{\Large \bf Gravitational Collapse of Baryonic and Dark matter}

\author{Dipanjan Dey} 
\email{dipanjandey.adm@charusat.edu.in}
\author{Pankaj S. Joshi}
\email{psjprovost@charusat.ac.in} 
\affiliation{International Center for Cosmology, Charusat University, Anand 388421, Gujarat, India}
\bigskip

\begin{abstract} 
A massive star undergoes a continual gravitational collapse when the pressures inside the collapsing star become insufficient to balance the pull of gravity. The Physics of gravitational collapse of stars is well studied. Using general relativistic techniques one can show that the final fate of such a catastrophic collapse can be a black hole or a naked singularity, depending upon the initial conditions of gravitational collapse. While stars are made of baryonic matter whose collapse is
well studied, there is good indirect evidence that another type of matter, known as dark matter, plays an important role in the formation of large-scale structures in the universe, such as galaxies. It is estimated that some eighty-five percent of the total matter in the universe is dark matter. Since the particle constituent of dark matter is not known yet, the gravitational collapse of dark matter is less explored. Here we consider first some basic properties of baryonic matter and dark matter collapse. Then we discuss the final fate of gravitational collapse for different types of matter fields and the nature of the singularity which can be formed as an endstate of gravitational collapse. We then present a general relativistic technique to form equilibrium configurations, and argue that this can be thought of as a general relativistic analog of the standard virialization process. We suggest 
a modification, where the Top-Hat collapse model of primordial dark matter halo formation is modified by using the general relativistic technique of equilibrium. We also explain why this type of collapse process is more likely to happen in the dark matter fields.

\end{abstract}
\maketitle

\section{Introduction}
Stars are born in the cloud of gas and dust which is known as nebulae. Nebulae are primarily made of hydrogen gas. Slightly overdense regions in that gas cloud become more dense due to their own gravitational pull. Now as hydrogen is baryonic matter, it slowly radiates energy as it is falling under its gravitational potential. During this time the radiation pressure is not enough to balance the gravitational pull. As time goes by, the gas becomes more dense and hotter. At last the time comes when the temperature becomes high enough to fuse two hydrogen atoms into a helium atom. This nuclear fusion releases a huge amount of radiation which resists the gas cloud to collapse further and a star is born. Stars shine as they burn their nuclear fuel mainly hydrogen, fusing it into helium and later other heavier elements. After shinning for millions of years, the time comes when the star runs out of the fuel and cannot resist its own gravitational pull, and it finally collapses under its own gravity.

The more massive a star is, the shorter the lifespan it has. For massive stars like 10 to 20 times the solar mass, they exhaust their nuclear 
fuel far quickly than our sun. For the sun-like stars, the hydrogen in the central region is used up first to make a core of helium, and then the hydrogen in the spherical shell around the core begins to fuse. This causes the star to grow gradually in size until the red giant phase is achieved. After that, the inner core collapses further and creates more heavy elements. The outer shells of star explode and create a planetary nebula. The inner core stops collapsing when the quantum pressure of degenerate electron gas balances the gravitational force. This central dense, compact object is known as a white dwarf. There is a mass limit of stars for which the final state of gravitational collapse is a white dwarf. The limit is approximately 1.4 times the solar mass. This mass limit was set by Subrahmanyan Chandrasekhar in 1934 \cite{Chandra1}. Stars having masses greater than that limit causes further collapse of the core. Generally, the outer layers of these massive stars explode as a supernova explosion. The collapsing core settles down to another equilibrium stage which is caused by quantum pressure of neutrons, and in this way, a neutron star is born \cite{Lieb},\cite{Latt},\cite{annu}. This neutron star is barely ten to twenty kilometers in size. For more massive stars having masses greater than about ten solar masses, the above-mentioned quantum pressures become insufficient to balance the gravity. Then the continual gravitational collapse of the star becomes inevitable.

Massive stars which have a mass of the order of about ten solar masses or more burn much faster, and they are far more luminous than our sun-like stars. The lifespan of these massive stars cannot be more than ten to twenty million years, whereas, for a sun-like star, the lifespan is in billions of years. So one of the most important questions in astronomy and astrophysics today is to investigate the final fate of these very short-lived massive stars. General relativity indicates that this type of continual gravitational collapse terminates into a space-time singularity if certain energy conditions hold throughout the collapse process. However, the presently available results so far in general relativity cannot predict the nature of such a singularity. One of the most intriguing problems in today's gravitational physics is whether the central ultra-high density region could be visible to outside observers, or these should always be covered by an event horizon of gravity. We will discuss this further in section (\ref{ccc}).

As we know, stars are made of baryonic matter. Besides gravitational interaction, baryonic matter particles can also interact with each other by means of strong, weak, and electromagnetic forces. Only with gravitational force, the stars would not form any kind of bound states. Due to these three interactions, the star can radiate away energy, and therefore it can go to the lower energy states in a gravitating system.
This process is responsible for the formation of stars, planets as well. Now, the situation becomes more interesting if we give our attention to larger scales, i.e. the galactic scales and the cosmological scale, where in fact the dark matter could dominate.

A typical sun-like star has a radius from about $10^5$ to $10^6$ km. 
The largest stars have radii of $10^{10}$ km approximately. On the other hand, the galactic scale ranges from the order of 10 to 100 kiloparsec (kpc), which is the scale of  $10^{16}$ km and higher. The cosmological scales start from 100 megaparsecs (mpc). The galaxy clusters scale falls in between the galactic scale and cosmological scale. 

In the galactic scale, it is observed that stars, rotating around the center of a galaxy, have an orbital velocity which remains constant or `flat', with increasing distance away from the galactic center \cite{Rubin},\cite{Rubin2}. This was a highly surprising result, as this kind of dynamics cannot be seen in a planetary system. In a planetary system, the velocities of planets decrease as the radial distance from the star increases. Flat galactic rotation curves indicate that each galaxy is surrounded by a significant amount of unknown matter which is
not actually seen, known as dark matter. It is estimated that the total dark matter mass should be around three times the total mass of baryonic mass in a galaxy. It is roughly considered that dark matter has a spherical shaped halo-like structure enshrouding each galaxy. It is named as "dark matter" because it does not appear to interact with the usual standard model particles, and therefore it is very difficult to detect a dark matter particle by the modern detectors. 

Therefore, Dark matter is supposed to interact with ordinary matter only through gravity. So, only by its gravitational effects, we can gather information, such as its positions, mass, density profiles, etc. The existence of dark matter also can be proved indirectly by the large-scale structure in the cosmological scale.  If there was no dark matter, then it can be shown that only baryonic matter would not lead to the present large-scale structure of the universe. Due to the inert property of dark-matter, the structures made by it can be distinguished from the structures made by baryonic matter. One could ask, how did the dark-matter start to form its structures. How did it form its halo-like 
structure around the galaxy? What are the internal properties of the dark matter halo? As we do not know what the dark matter is, these are among the most difficult questions of cosmology.

In its cosmological scale, our universe is almost flat and homogeneous.  In that scale, one can explain the dynamics of our universe by spatially flat Friedmann-Lemaitre-Robertson-Walker (FLRW) metric. Perturbations in that metric are seeded by the matter density perturbations of our universe. The quantum fluctuations during the inflationary era are responsible for such density perturbations. During the inflationary era, the universe expanded exponentially, but the Hubble horizon remained constant. Therefore, quantum perturbation modes during the inflationary era left the Hubble horizon and reentered the horizon after the inflationary era. The perturbation modes became classical before they reentered the horizon. Therefore, the perturbation modes which enter the Hubble horizon at the end stage of radiation domination create density perturbation in matter \cite{Hogan:1985bc},\cite{lidd} \cite{sengor},\cite{white}. 

As the baryonic matter decoupled after dark matter decoupling, the perturbations in dark-matter started to grow before the growth of perturbations in baryonic matter field. The inert property of dark matter helps it to decouple first and form first primordial structures in our universe. The dynamics of density perturbations in dark matter can be explained by linear perturbation theory as long as the density contrast $\frac{\delta\rho}{\bar{\rho}}<1$, where $\delta\rho=\rho-\bar{\rho}$ and $\bar{\rho}$, $\rho$ are the background density and total density of perturbed area respectively. However, when this density contrast becomes almost one, linear perturbation theory is insufficient to explain the dynamics of the perturbation modes. That is the time when perturbation modes entered into the non-linear regime and form overdense patches in our universe. As we have mentioned, in the cosmological scale we use spatially flat FLRW metric to describe the dynamics of homogeneous, isotropic and expanding universe. Therefore, the background of the overdense regions is expanding. It is generally considered that as the background FLRW universe expands in a homogeneous and isotropic fashion, these overdense matter distributions also expand initially in an isotropic and homogeneous fashion, but with a slower rate. As the density contrast of the overdense regions increases with time, they gradually detach from the cosmic expansion. As we have mentioned earlier, dark matter forms its structure before the formation of baryonic structures. Therefore, these overdense patches can be thought of as a very early stage of primordial dark-matter halo. Initially, these hallo-like structures expand with the background, then the time comes when it stops expanding and starts collapsing due to its own gravitational pull \cite{Frenk},\cite{Cooray}. 

As we have discussed, except gravity, baryonic matter can interact with each other by the extra three forces. Therefore, they can form different types of complicated celestial objects. On the other hand, dark matter does not show any observational evidence of interaction with any baryonic matter, except gravity. Therefore, it is unable to radiate energy. As it cannot radiate energy, it cannot form any bound objects at stellar scale, like stars, planets, etc. Dark matter can form structures at galactic scale or higher.  As we do
not know the particle constituent of dark matter, it is very difficult to predict the final fate of dark matter collapse. Generally, dark matter is considered as collisionless cold matter. Here cold matter means dark matter particle has velocity much less than the velocity of light. This is known as cold dark matter (CDM) model \cite{DelPo}. This model achieves great success in explaining how the large-scale structure of the universe evolves with time. In cosmological scales, this model has excellent agreement with observational results. However, this model is unable to solve some problems in the galactic scale. For example, it does not have a satisfactory explanation of why the center of the galaxy is core-like, why it is not cuspy at the center. There are also other problems like: `missing satellite' problem, `too big to fail' problem, and others. 

There is a modified version of CDM, and it is known as the $\Lambda$ CDM \cite{DelPo}, where $\Lambda$ stands for the cosmological constant. To incorporate the accelerated expansion of the universe in the CDM model this modified version has been proposed. However, the problems in the galactic scale remain unsolved \cite{Bullock},\cite{Weinberg}. To solve these problems, self-interacting dark matter (SIDM) model is proposed \cite{Spergel}. In this model, the dark matter particles are considered to be self-interacting, which cannot be neglected. This model has more promising  
solutions for those problems in galactic scale. 

All these models have a common assumption for the equilibrium state of collapsing dark matter halo. They all use the so-called virialization technique to explain the final equilibrium state of a dark matter collapse \cite{Merr}. The Virial theorem generally gives a relation between the average kinetic energy of a stable system consisting of $N$ number of particles, which is bound by some potential energy, with that of the total potential energy. A system is said to be virialized when the particles of the system have some dynamics or kinetic energy, but overall the system is static. It can be mathematically proved that virialization should occur after a very large time only.  When the particles in a gravitating system have only gravitational interaction between each other, then at virialized state the system should follow,
\begin{equation}
\langle T\rangle_{\tau\rightarrow\infty} = -\frac12\Bigg\langle \sum_{i=1}^{N}{\bf F}_i\cdot{\bf r}_i  \Bigg\rangle_{\tau\rightarrow\infty}\, ,
\label{vir1}
\end{equation}
where ${\bf F}_i$ is net the force on the $i$th particle, ${\bf r}_i$ is the position vector of $i$th particle and $T$ is the total kinetic energy of the system.
Here, the time averaging is done over time $\tau$. The time $\tau$ should be very large so that the system can achieve the virialized state.
For a gravitating system we can write,
\begin{equation}
\sum_{i=1}^{N}{\bf F}_i\cdot{\bf r}_i = -V_T\, ,
\label{vir2}
\end{equation}
where $V_T$ is the total potential energy of the system. Using eqs.~(\ref{vir1}), (\ref{vir2}) we finally get the virialization equation,
\begin{equation}
\langle T\rangle=-\frac12 \langle V_T \rangle\, .
\end{equation}
For the CDM case, system virializes through different processes: phase mixing, violent relaxation \cite{Lynd}, chaotic mixing, and the Landau damping \cite{Merr}. For SIDM case a system virializes through more complicated mechanism which we will not discuss here \cite{tulin},\cite{Saxton}. As we have discussed above, dark matter forms first structures in the universe. When the Baryonic matter cools down, it starts forming its structure due to the gravitational potential of overdense primordial dark matter halos \cite{Birnboim:2003xa}, \cite{jing}, \cite{Silk}, \cite{tanu}. 
The virialization process, which we have discussed above, is a classical Newtonian process. Therefore, it does not incorporate any general relativistic phenomena during the collapse, e.g. the apparent horizon, singularity, black hole, event horizon, etc. One could, however, ask, if the dark matter is capable of creating these types of non-trivial celestial objects. As we discussed earlier, we do not know the particle constituent of dark matter, so it is very difficult to answer this question. 

This paper is organized as follows. In section (\ref{ccc})   we will discuss the causal structure of collapsing space-time. In that section, we will discuss the possible nature of singularity which can be formed as an end state of gravitational collapse. In section (\ref{dust}), the final fate of the gravitational collapse of homogeneous and inhomogeneous dust will be discussed. In section (\ref{stable}) we will review a work \cite{Joshi:2011zm}, where it is shown how the presence of pressure in a collapsing system can equilibrate the system to a stable configuration. In section     (\ref{Halo}) we will review another work \cite{Bhatt}, where it is shown that without using virialization technique one can describe primordial dark matter halo formation by using a general relativistic technique of equilibrium. In the last section, we summarize and
discuss some future outlook.

\section{Stellar Collapse and Cosmic Censorship Conjecture}
\label{ccc}
In General relativity, it is inevitable to have a space-time singularity as the end state of a gravitational collapse for a physically realistic matter cloud, that maintains certain positivity of energy conditions. At the singularity, a space-time is timelike or null geodesically incomplete. In a singular space-time, it is possible for at least one freely falling particle or photon to end its existence within a finite time (or affine parameter) or to have begun its existence a finite time ago. The singularity theorems \cite{Seno},\cite{Hawk} tell us that the gravitational collapse of a massive collapsing matter cloud would terminate into a space-time singularity if the following conditions are satisfied,

\begin{itemize}
\item Throughout the collapse process the following energy condition should hold,
\begin{equation}
R_{ab}X^aX^b\geq 0\,\, ,
\label{strong1}
\end{equation}
where $X^a$ is a non-spacelike geodesic and $R_{ab}$ is the Ricci tensor. For timelike geodesics, the above condition is known as {\it strong energy condition} \cite{poiss}. The strong energy condition states that gravity should be always attractive. Now, if the space-time curvature is seeded by a perfect fluid, then this condition can be written as,
 \begin{eqnarray}
 \rho+3P\geq 0\, \, ,\,\, \rho+P\geq 0\,\, .
 \label{strong2}
 \end{eqnarray}
 Now, for a light-like geodesic, the above condition (eq.~(\ref{strong1})) is known as the {\it null energy condition}. For a perfect fluid, this can be written as,
 \begin{eqnarray}
 \rho+P\geq 0\, \, .
 \label{null1}
 \end{eqnarray}
From eqs.~(\ref{strong2}), (\ref{null1}), it can be seen that the validity of strong energy condition throughout the collapse implies the validity of null energy condition. 
%weak energy condition should be valid. Weak energy condition states that energy density should be always positive with respect to any observer. This energy condition mathematically can be written as,
% \begin{eqnarray}
% \rho >0, \, \, \rho+P_r\geq 0,\, \, \rho+P_\perp\geq 0,\, \, ,
 %\end{eqnarray}
% where $P_r$ and $P_\perp$ are the radial and azimuthal pressure of fluid.

\item The causal structure of the collapsing matter should obey strong causality condition. There should not be any closed timelike geodesics in the space-time.

\item There should be a closed future trapped surface in the collapsing matter cloud.

\end{itemize}

The final space-time becomes timelike or null geodesically incomplete when the above conditions are fulfilled. From the above discussion, it should be noted that some situations can arise where space-time is null geodesically incomplete but timelike geodesically complete, e.g., the Reissner-Nordstrom solution. Here, we should also make some comments on the weak energy condition as it is considered as a necessary condition for a realistic matter field to satisfy. The weak energy condition states that energy density should always be positive with respect to any observer. Mathematically it can be written as,
\begin{equation}
T_{ab}u^au^b\geq 0\,\, ,
\label{weak1}
\end{equation}
where $u^a$ is any timelike vector.
For a perfect fluid case, one can write the above equation as,
 \begin{eqnarray}
\rho >0, \, \, \rho+P\geq 0,\, \, .
\label{weak2}
\end{eqnarray}
Therefore, weak energy condition implies null energy condition but does not imply strong energy condition. Weak energy condition is considered to be valid for any regular point in a space-time. On the other hand, strong energy condition can be violated for some circumstances. For the vacuum energy dominated universe (e.g. inflationary era), and for the dark-energy-dominated universe, strong energy condition is violated. However, for those cases also the weak energy condition is always valid.

So, the singularity theorems tell us on the conditions to have possible singular space-times as the end state of gravitational collapse. However, the key point is, these theorems are unable to predict whether the singularity that occurs can be visible to an asymptotic observer. It is generally assumed that if the final state of a catastrophic continual gravitational collapse terminates into a singular space-time, then the final singularity should always be covered by a null hypersurface, which is known as the event horizon. On this null 
hypersurface, the incoming null geodesics are converging and outgoing null geodesics are parallel,
that is, they have a vanishing expansion. The singularity inside this null hypersurface is known as spacelike singularity. As timelike or null geodesics are converging inside the event horizon, the three-dimensional volume, which is covered by the event horizon, becomes totally black or invisible to any outside observer. Therefore, this region is called a black hole. Any information about the singularity inside the black hole cannot reach to the outside observer. 

As was stated above, it was generally considered that gravitational collapse of any realistic matter field always terminates into a black hole.    However, there is no mathematical interpretation or proof behind this assumption. So this is a conjecture.  
In 1969 Roger Penrose gave this conjecture \cite{pen1}, \cite{Pen2}, which states that continual gravitational collapse of any physically realistic matter field will generically (i.e. stable under small perturbations) terminate into a black hole if certain energy conditions are obeyed throughout the collapse. So this conjecture states that every space-time singularity should be covered by an event horizon of gravity. That it should not be globally naked, visible to faraway observers, is known as the Weak Cosmic Censorship Conjecture (WCC). In 1978 Penrose suggested a stronger version of this conjecture which states that any singularities that arise from regular initial data are not even locally visible. This is known as Strong Cosmic Censorship Conjecture (SCC). 

\begin{figure}[t]
\includegraphics[scale=0.80]{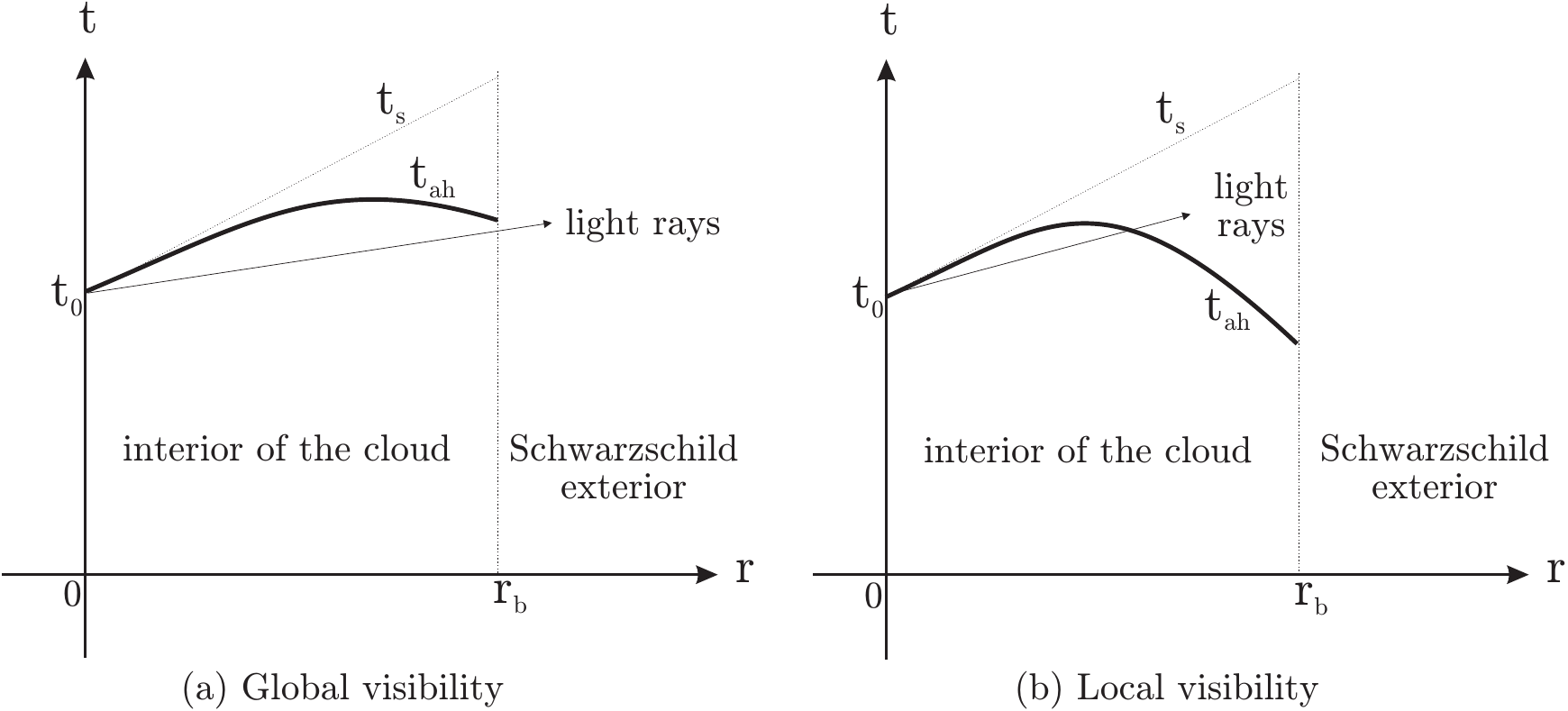}
\caption{(a) Here, light rays from the central singularity can escape the collapsing matter cloud before the apparent horizon formation, and they reach  the asymptotic observer. Therefore the central singularity can be visible for faraway observers.
(b) Here, light rays fall into the apparent horizon before they reach the boundary of collapsing matter cloud. However, as these propagate inside the cloud, the singularity is locally visible. The picture is taken from \cite{Joshi:2012mk}.  }
\label{visibility}
\end{figure}

To verify these two conjectures, we have to compare the time of singularity formation and the time of apparent horizon or trapped surfaces formation. Trapped surfaces are the two-dimensional hypersurfaces on which both the incoming and outgoing null geodesics are converging \cite{Bizon:1988vv}, \cite{Ellis}, \cite{Weinberg:1972kfs}. This type of 2-surfaces can form inside a collapsing matter cloud. The apparent horizon is the boundary of these trapped surfaces. A singularity is globally visible when null geodesics starting from space-time singularity can escape the collapsing matter before the formation of the apparent horizon, and then reaches the asymptotic observer. A singularity is locally visible when space-time singularity at the center of the collapsing matter is formed before the formation of apparent horizon around the center so that singularity could be visible locally, but it is not globally visible. The global and local visibility of singularity is described diagrammatically in fig.~(\ref{visibility}).  SCC actually avoids any kind of timelike or null singularities. As this conjecture is yet to be proven, it is one of the most  challenging unsolved problems of gravitational physics.

From the above discussion, it is clear that the position and formation of apparent horizon plays a very important role in the causal structure of a collapsing space-time. The position of apparent horizon can be derived from the expression of the expansion parameter ($\theta$) of null geodesic congruences. As an example, one can examine the position of apparent horizon in the following LTB metric,
\begin{eqnarray}
 ds^2=  -dt^2 +\frac{R^{\prime 2}(r,t)dr^2}{1+E(r)}+R^2(r,t)d\Omega^2 \, ,\nonumber
 \label{LTBMetric}
\end{eqnarray}
where $R(t,r)$ is the physical radius and ($t,r$) are co-moving time and radius respectively and $E(r)$ is a real valued function of $r$. Here and in general a prime over a function denotes the partial derivative with respect to $r$ and a dot over a function denotes the partial derivative with respect to $t$. Now for this space-time one can derive the following expression of expansion parameter for outgoing radial null geodesics,
\begin{equation}
\Theta_l=\frac{2}{R}\left(\dot{R}+\sqrt{1+E(r)}\right)\, .
\label{expanout}
\end{equation}
Similarly, one can get the following expression of $\Theta_n$ for incoming radial null geodesics,
\begin{equation}
\Theta_n=\frac{2}{R}\left(\dot{R}-\sqrt{1+E(r)}\right)\, .
\label{expanin}
\end{equation}
One can express $\dot{R}$ in term of Misner-Sharp mass term ($F$)  \cite{Misner}, \cite{May:1966zz}. Mathematically the Misner-Sharp mass  can be written as,
\begin{equation}
F(r,t)=R(r,t)\left(-E(r)+\dot{R}(r,t)^2\right)\, .
\label{Misner}
\end{equation} 
So, during collapse one can write the expression of $\dot{R}$ as,
\begin{equation}
\dot{R}=-\sqrt{\frac{F}{R}+E(r)}\, ,
\end{equation}
where the negative sign indicates that matter is collapsing. Using this expression of $\dot{R}$, we get the following expressions of expansion parameter for outgoing and incoming radial null geodesics,
\begin{equation}
\Theta_l=\frac{2}{R}\left(\sqrt{1+E(r)}-\sqrt{\frac{F}{R}+E(r)}\right)\, ,\,\,\,\, \Theta_n=-\frac{2}{R}\left(\sqrt{1+E(r)}+\sqrt{\frac{F}{R}+E(r)}\right)\,\, .
\label{LTBtheta}
\end{equation}
The above equations indicate that the expansion parameter for incoming radial null geodesics should always be less than zero. On the other hand, the expansion parameter for outgoing radial null geodesics could change its sign. For $R=F$, $\Theta_l$ becomes zero, for $R>F$, $\Theta_l$ becomes negative and for $R<F$, $\Theta_l$ becomes positive. Now, it is known that the expansion parameter for outgoing radial null geodesics becomes zero at the apparent horizon. Mathematically, apparent horizon can be expressed (\cite{Lasky:2006hq},\cite{Lasky:2006mg}) as the surface on which,
\begin{equation}
\Theta_l=0\,\,,\, \Theta_n<0\, .
\label{apparent}
\end{equation}
From eqs.~(\ref{LTBtheta}),(\ref{apparent}) one can say that the condition for existence of apparent horizon is,  
\begin{equation}
F(r_c)=R(r_c,t_{AH})\, .
\label{conApp2}
\end{equation}
The above condition states that when at a particular comoving radius $r_c$ and comoving time $t_{AH}$ the corresponding $R(r_c,t_{AH})$ reaches the value of $F(r_c)$, an apparent horizon will form in LTB space-time.
So, throughout the collapse, the condition to avoid the apparent horizon for any co-moving radius $r$ is,
\begin{equation}
\frac{F(r,t)}{R(r,t)}<1\, .
\label{app1}
\end{equation}

We note that global and local visibility of singularity is a very big problem to investigate in gravitation theory. In many literature available \cite{Vaz:1995ig}-\cite{Joshi:2007am}, \cite{Adler:2005vn}, over the last three decades, this issue has been extensively investigated. The results show that there is always a possibility that singularity becomes globally and locally visible, at least for some time. The genericity of these naked singularities was also investigated \cite{Joshi:2011qq},\cite{Satin:2014ima} \cite{Joshi:2007hq}, and it was found that there can be generic naked singularity solutions of Einstein equation.   This does not really disprove the CCC, as there is always a debate on the physical viability of the initial conditions which are taken for a collapsing scenario. However, this also implies that the cosmic censorship, if at all valid, can hold only under very fine-tuned conditions, and is thus severely restricted in any case.  

%\begin{figure}[t]
%\includegraphics[scale=0.70]{visibility.eps}
%\caption{(a) Here light rays, from central singularity, can escape the collapsing matter cloud before the apparent horizon formation and reach to the asymptotic observer. Therefore the central singularity can be visible from outside.
%(b) Here light rays fall into the apparent horizon before it reach the boundary of collapsing matter cloud. However, as it propagate inside the cloud, it is locally visible.  }
%\label{visibility}
%\end{figure}

\section{Gravitational Collapse Models}
\label{dust}
In 1939 Oppenheimer, Snyder \cite{Oppen} and Datt \cite{Datt} gave a continual gravitational collapse model . This is homogeneous (energy density independent of spatial coordinates), spherical collapse of a pressure-less fluid (dust). This is the simplest model of continual gravitational collapse of any spherically symmetric matter field. When a star having tens of times of the solar masses, exhausts its internal nuclear fuel and goes into catastrophic gravitational collapse, it inevitably forms a space-time singularity as the final state of gravitational collapse. This type of collapse process has to be   modeled using general relativity. While OSD was an attempt in that direction, as we know, any physically realistic collapsing matter has inhomogeneity in density during the collapse, and will also have non-zero internal pressures. Therefore the OSD collapse can only give a
rather idealized model of the continual gravitational collapse process.  
This was the first general relativistic model of continual gravitational collapse which predicted a black hole as an end state of collapse.

Before we discuss the final fate of the OSD collapse, we should mention the regularity conditions for a physically realistic collapsing scenario. A collapsing system should follow the following regularity conditions,
\begin{itemize}
\item During the collapse every space-time point of the collapsing system must be regular so that every physical parameter of that space-time should have finite value throughout the collapse.
\item Weak energy condition, which we have discussed earlier, should be maintained throughout the collapse.
\item Throughout the collapse there should not be any shell-crossing singularity \cite{
Hellaby:1985zz},\cite{szekeres1999shell},\cite{Joshi:2012ak}. The partial derivative of the physical radius $R$ with respect to comoving coordinate $r$ should always be greater than zero to avoid a shell- crossing singularity:
\begin{equation}
R'(t,r)>0.
\label{shell1}
\end{equation}
\end{itemize}
Keeping all these collapse regularity conditions in mind, we can now start to discuss the final fate of OSD collapse, which is a spherically symmetric, homogeneous, dust collapse.

%\begin{figure}[t]
%\centerline{\includegraphics[width=9cm]{bh.eps}}
%\caption{Oppenheimer-Snyder-Datt (OSD) collapse: The figure is showing dynamical evolution of a spherically symmetric, homogeneous dust cloud collapse. Here both the strong and weak cosmic censorship conjectures are valid. \label{f:one}}
%\end{figure}

\subsection{Dust Collapse with Homogeneous Density}
The OSD collapse can be mathematically described by spherically symmetric Lemaitre-Tolman-Bondi (LTB) metric \cite{LTB1}, \cite{LTB2}, \cite{LTB3},
\begin{eqnarray}
 ds^2=  -dt^2 +\frac{R^{\prime 2}(r,t)dr^2}{1+E(r)}+R^2(r,t)d\Omega^2 \, .\nonumber
 \label{LTBMetric}
\end{eqnarray}
As said earlier, $R(t,r)$ is the physical radius and ($t,r$) are comoving time and radius. From Einstein equations, one can get,
\begin{eqnarray}
  \rho=\frac{F^\prime}{R^2 R^\prime}\,,\,\,\,\,
  P_r =P_\perp =-\frac{\dot{F}}{R^2 \dot{R}}=0\,,\,\,\,\,
\label{p0}
\end{eqnarray}
where $\rho$, $P_r$ and $P_\perp$ are the energy density, radial pressure and azimuthal pressure respectively. 
\begin{figure}[h!]
\centerline{\includegraphics[width=9cm]{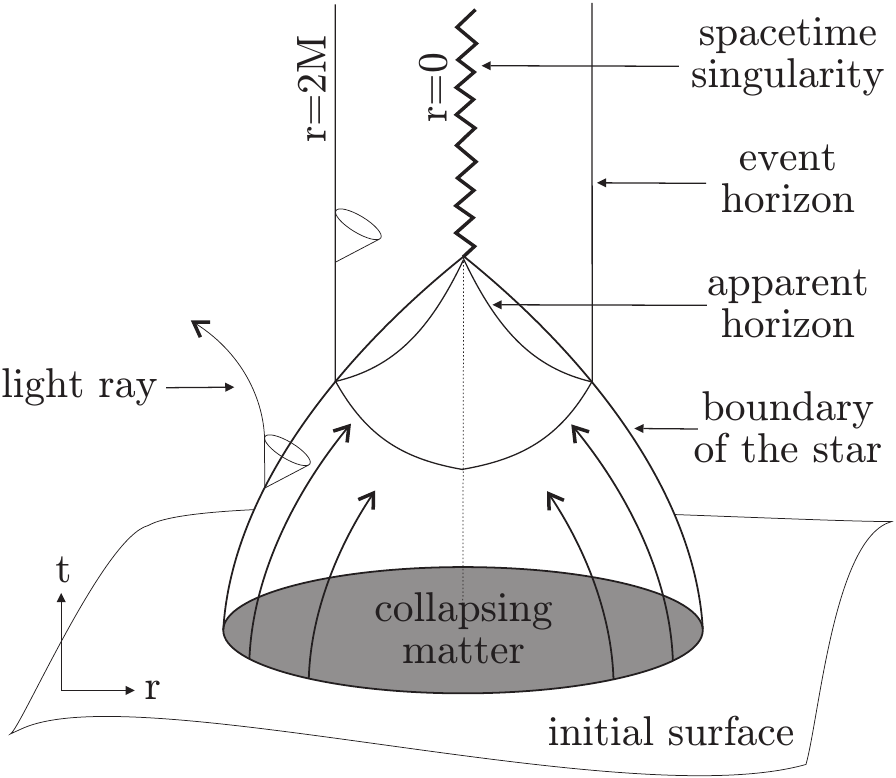}}
\caption{Oppenheimer-Snyder-Datt (OSD) collapse: The figure is showing dynamical evolution of a spherically symmetric, homogeneous dust cloud collapse. Here, both the strong and weak cosmic censorship conjectures are valid. The picture is taken from \cite{Joshi:2012mk}. \label{f:one}}
\end{figure}
The expression  of the Misner-Sharp mass $F$ \cite{Misner}  is given in eq.~(\ref{Misner}). Using the $\rho$ expression in eq.~(\ref{p0}), we can write,
\begin{equation}
F(t_i,r)=\int_0^r\rho(t_i,\tilde{r})\tilde{r}^2d\tilde{r}\,\, ,
\label{F1}
\end{equation}
where we consider $R(r,t_i)=r$ and $t_i$ is the initial time.
For homogeneous gravitational collapse, we have the mass function $F$ given as,
\begin{equation}
F(r,t_i)=F_0 r^3\,\, ,
\label{F2}
\end{equation}
where $F_0=\frac{\rho(t_i)}{3}$.
From eq.~(\ref{p0}) we can see that for a dustlike collapse, $F$ becomes time-independent. Therefore, the expression of $F$ in eq.~(\ref{F2}) must remain unchanged throughout the collapse. Now, in the LTB metric we always have a scaling degree of freedom. Therefore we can consider the physical radius as, 
\begin{equation}
R(t,r)=rf(t,r)\,\, ,
\label{R1}
\end{equation}
with $f(t_i,r)=1$. For a collapse process we should always have $\dot{f}(t,r)<0$. In a collapse process 
$f$ always decreases with time monotonically. Therefore, we can use $f$ as `time-coordinate` replacing $t$ and every function of $t$ can be thought of as the function of $f$.  The expression of Misner-sharp mass in eq.~(\ref{Misner}) is the equation of motion of this system. Using that equation one can write time $t$ as a function of $r$ and $f$:
\begin{equation}
t(f,r)=t_i+\int^1_f\frac{d\tilde{f}}{\sqrt{\frac{F_0}{\tilde{f}}+\frac{E(r)}{r^2}}}\,\, .
\label{texp}
\end{equation}
As during the collapse $E(r)$ should be regular at the center, we can write $E(r)$ as,
\begin{eqnarray}
E(r)&=& r^2b(r)\nonumber\\
    &=& r^2\left(b_0+b_1 r+b_2 r^2+......\right)\,\, ,
\label{E1}    
\end{eqnarray}
where we have expanded $b(r)$ in a Taylor series around the center. Therefore, The coefficients $b_1, b_2,...$ can be written as $b_1=\frac{db}{dr}\rvert_{r=0}$,
$b_2=\frac{1}{2}\frac{d^2b}{dr^2}\rvert_{r=0}$ etc., close to the center. Using the eq.~(\ref{E1}), we can now rewrite the eq.~(\ref{texp}) as,
\begin{equation}
t(f,r)=t_i+\int^1_f\frac{d\tilde{f}}{\sqrt{\frac{F_0}{\tilde{f}}+b(r)}}\,\, .
\label{texp2}
\end{equation}
Now, the condition for singularity is $$R(t_s,r)=0\,\, ,$$
where $t_s$ is the comoving time of singularity formation for the comoving radius $r$. Therefore, at the time of singularity formation we can write,
\begin{equation}
f(t_s,r)=f_s=0\,\, .
\label{singf}
\end{equation}
Using eq.~(\ref{texp2}) and eq.~(\ref{singf}), we can write the time of singularity formation as,
\begin{equation}
t_s(r)=t_s(0,r)=t_i+\int^1_0\frac{d\tilde{f}}{\sqrt{\frac{F_0}{\tilde{f}}+b(r)}}\,\, .
\label{tsing1}
\end{equation}
For simplicity we can assume, $b(r)=0$, which is called a marginally bound collapse. With that assumption we can now fully integrate the above equation and get the analytic expression of the time of singularity formation:
\begin{equation}
t_s=t_i+\frac23\frac{1}{\sqrt{F_0}}
\label{tsing2}
\end{equation}
The above expression of $t_s$ indicates that every portion of the collapsing cloud shrinks into a point simultaneously. 
The condition for apparent horizon is $R(r,t_{AH})=F(r)$, where $t_{AH}$ is the comoving time of apparent horizon formation at comoving radius $r$. Using the eq.~(\ref{texp2}) we can write the time $t_{AH}$ of apparent horizon formation as, 
\begin{eqnarray}
t(f_{AH},r)=t_{AH}(r)&=&t_i+\int_{f_{AH}}^1\frac{d\tilde{f}}{\sqrt{\frac{F_0}{\tilde{f}}+b(r)}}\,\, \nonumber\\
&=&t_i+\int_{0}^1\frac{d\tilde{f}}{\sqrt{\frac{F_0}{\tilde{f}}+b(r)}}-\int^{f_{AH}}_0\frac{d\tilde{f}}{\sqrt{\frac{F_0}{\tilde{f}}+b(r)}}\nonumber\\
&=&t_s-\int^{f_{AH}}_0\frac{d\tilde{f}}{\sqrt{\frac{F_0}{\tilde{f}}+b(r)}}.
\label{texp3}
\end{eqnarray}
Now, from the condition of apparent horizon formation, we get: $f_{AH}=r^2F_0$. Finally, using eq.~(\ref{texp3}) and assuming $b(r)=0$ we get,
\begin{equation}
t_{AH}(r)=t_s-\frac23 F_0 r^3\,\, .
\label{eqahosd}
\end{equation}
 As $F_{0}>0$, for this case we always have  $t_{AH}<t_s$ which means, for homogeneous dust collapse the singularity cannot be locally naked and as it is not locally naked it is not globally naked also. Here we have shown the results only for the case where $b(r)=0$. However, it can be shown that for any regular value of $b(r)$, we always get $t_{AH}<t_s$. In fig.~(\ref{f:one}), a homogeneous, spherical dust collapse is shown diagrammatically. In that diagram one can see how apparent horizon forms before the singularity formation at $r=0$. Therefore, for spherically symmetric, homogeneous dust collapse, it is inevitable to have a singularity as a final state of gravitational collapse, and that singularity is not visible locally
or globally.

%\begin{figure}[t]
%\centerline{\includegraphics[width=9cm]{ns.eps}}
%\caption{Inhomogeneous Dust Collapse: A space-time singularity, which is formed from an inhomogeneous, spherically symmetric dust collapse, can be visible to external observers in the universe,
%in violation to both the weak and strong cosmic censorship conjecture. \label{f:two}}
%\end{figure}

\begin{figure}[t]
\centerline{\includegraphics[width=9cm]{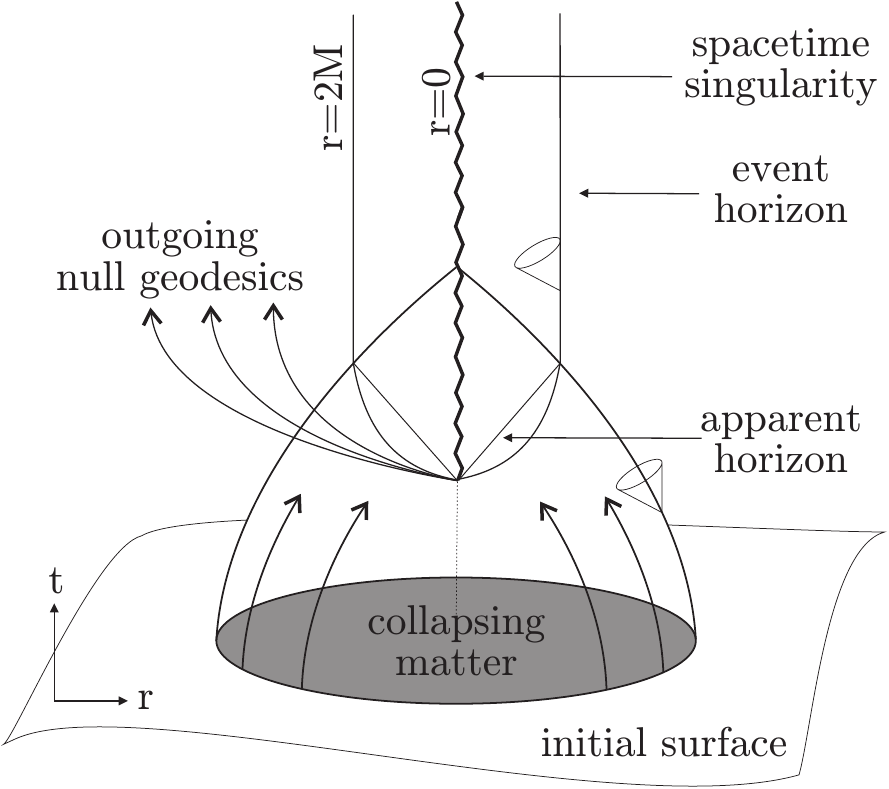}}
\caption{Inhomogeneous Dust Collapse: A space-time singularity, which is formed from an inhomogeneous, spherically symmetric dust collapse, can be visible to external observers in the universe,
in violation to both the weak and strong cosmic censorship conjectures. The picture is taken from \cite{Joshi:2012mk}. \label{f:two}}
\end{figure}

\subsection{Dust Collapse with Inhomogeneous Density}
Collapse of an inhomogeneous dust cloud is more realistic than the collapse of a totally homogeneous dust cloud \cite{Bondi:1947fta},\cite{joshibook}. In any physically realistic system, which is collapsing under its own gravitational pull, the density generally decreases as the radial distance from center increases. One can consider a simple model case where,
\begin{equation}
\rho(t_i,r)=\rho_0+\rho_{2}r^2\,\, ,
\end{equation}
where $\rho_{2}<0$ . For this energy density we have,
\begin{equation}
F(r)=F_{0}r^3+F_{2}r^5\nonumber\,\, ,
\end{equation}
where $F_{2}<0$. Using the same technique, which was discussed above, we get the following expression of $t_s(r)$ and $t_{AH}(r)$ for inhomogeneous dust collapse,
\begin{eqnarray}
\label{ts2}
t_{s}(r)&=&t_s(0)-\frac{F_2}{3F_{0}^{\frac32}}r^2+....\,\, ,\\
t_{AH}(r)&=&
t_s(r)-\frac23F_{0}r^3=t_s(0)-\frac{F_2}{3F_{0}^{\frac32}}r^2-\frac23F_{0}r^3+ ....\,\, ,
\label{tah2}
\end{eqnarray}
where $t_s(0)$ is the time of singularity formation for the center at $r=0$ and it can be written as,
\begin{equation}
t_s(0)=t_i+\frac23\frac{1}{\sqrt{F_0}}\,\, .
\label{tsing2}
\end{equation}
From eq.~(\ref{ts2}) one can see that in an inhomogeneous dust collapse, the time of singularity  formation is different for different values of $r$.
It can also be seen that around the center the $r^2$ term will dominate over $r^3$ term and as $F_2<0$ we have:
$$t_{AH}(r)>t_s(0)\,\,.$$ So for this case, the singularity must be at-least locally naked. So the SCC is violated here. For a general form of $b(r)$ one can show that singularity can also be globally naked which is a violation of WCC \cite{Joshi:2004tb}, \cite{Joshi:1993zg}. In fig.~(\ref{f:two}), it is shown how naked singularity, which is formed as an end state of inhomogeneous spherical dust collapse, can be globally visible. So it is thus seen that an inhomogeneity in density can change the nature of singularity. 

From the above discussion, we can conclude that any spherically symmetric homogeneous or inhomogeneous dust collapse should terminate into a black hole or naked singularity, in a finite amount of proper time, depending on the initial density profile for the matter cloud. We have not shown or discussed here that these collapsing scenarios do terminate into a stable configuration as a final stage of gravitational collapse. However, in general relativity, it is possible to have collapsing solutions which can terminate into a stable equilibrium singular or regular space-time as the end state of gravitational collapse. This technique has been developed and can be found in many literature \cite{Cooperstock:1996yv} - \cite{Joshi:2011zm}. 
We now consider these solutions briefly in the next section, 
towards a possible application to collapse for larger scales other
than stellar collapse.

\section{Stable Configuration from Gravitational Collapse }
\label{stable}
In the previous section, we have seen that the inclusion of inhomogeneity can change the nature of singularity. In this section, we will see how the pressure inside a collapsing matter field can possibly form stable configurations at the end stage of gravitational collapse.
To describe this type of gravitational collapse we have to start with a general collapsing space-time. 
% Then counting the number of equations and unknown functions, one can calculate the number of degrees of freedom. With the help of these degrees of freedom and considering some initial conditions, one can see that a collapsing system can achieve certain equilibrium state maintaining some collapsing conditions at the final state of collapse. 
The metric of a general spherical collapsing system can be written as,
\begin{eqnarray}
ds^2 = - e^{2\nu(r,t)} dt^2 + {R'^2\over G(r,t)}dr^2 + R^2(r,t) d\Omega^2\, ,
\label{genmetric}
\end{eqnarray}
where $\nu(r,t)$, $R(r,t)$ and $G(r,t)$ are real functions of $r$ and $t$. The above metric can describe a spherical symmetric
gravitational collapse as those functions are independent of azimuthal coordinates. One can describe a spherically symmetric, inhomogeneous, non-dust like gravitational collapse using this metric maintaining some collapsing conditions which we have discussed before. The collapsing conditions and the initial conditions of the collapse will give some constrains on the functions of above metric which should be maintained throughout the collapse. Like LTB metric, here also we have a scaling degree of freedom due to which we can write
\begin{eqnarray}
R(r,t)= r f(r,t)\,,
\label{scaling1}
\end{eqnarray}
where $f(r,t)$ is a real and positive valued function of $r,t$.
Now, for the present metric the Misner-Sharp mass term can be written as,
\begin{eqnarray}
F = R\left(1 - G + e^{-2\nu} \dot{R}^2\right)\,.
\label{ms}  
\end{eqnarray}
If the general collapsing space-time is seeded by an anisotropic fluid, one can write the energy density and radial pressure as,
\begin{eqnarray}
  -T^0_{\,\,\,\,0}=\rho=\frac{F^\prime}{R^2 R^\prime}\,,\,\,\,\,
  T^1_{\,\,\,\,1}=P_r = -\frac{\dot{F}}{R^2 \dot{R}}\,,
\label{eineqns}
\end{eqnarray}
and the expression of azimuthal pressures can be derived from $G_2^2$ term and $T^{\mu\nu};\nu=0$ equation. Using the $G_2^2$ and $T^{\mu\nu};\nu=0$ one can get an equation which actually relates  $\rho, p_{r}, p_{\perp}, \nu^{\prime}$ and the equation is,
\begin{equation}
    \nu'=2\frac{P_\perp-P_r}{\rho+P_r}\frac{R'}{R}-\frac{P_r'}
{\rho+P_r} \,\, .
\label{pthecon}
\end{equation} 
The vanishing of the non-diagonal terms in the Einstein
equations gives,
\begin{eqnarray}
\dot{G}=2\frac{\nu^\prime}{R^\prime}\dot{R}G\,.
\label{vanond}
\end{eqnarray}
For an isotropic energy-momentum tensor where
$P_r=P_\perp$, the Einstein equations give,
\begin{eqnarray}\label{P_rho}
  \rho = {F'\over R^2 R'}\;, \,\; P_r = P_\perp=- {\dot F\over R^2\dot R} ~,
\label{P_rho}
\end{eqnarray}
while eq.~(\ref{vanond}) remains the same for an isotropic fluid and eq.~(\ref{pthecon}) becomes,
\begin{equation}
   \nu'=-\frac{P_r'}
{\rho+P_r}  \,\, .
\end{equation}
Now, we are going to discuss the final state of gravitational collapse of an anisotropic fluid.
For simplicity, we can consider zero radial pressure and non-zero azimuthal pressure. For this case, we have 4 equations,
\begin{eqnarray}
  \rho=\frac{F^\prime}{R^2 R^\prime}\,,\,\,\,\,
   P_\perp = \frac12 \rho R \frac{\nu^\prime}{R^\prime}\,,\,\,\,\,
  \dot{G}=2\frac{\nu^\prime}{R^\prime}\dot{R}G\,,\,\,\,\,F = R\left(1 - G + e^{-2\nu} \dot{R}^2\right)\,,
\label{eineqns2}
\end{eqnarray}
and 6 unknown functions ($\rho, P_{\perp}, R, G, F, \nu$ ), so we always have the freedom to choose the functional expressions of two free functions.
As we have discussed previously, for zero radial pressure, the Misner-Sharp mass term becomes time-independent. 
Let us consider the initial physical radius as,
\begin{equation}
R(t_i,r)=rf(t_i,r)=r\,\,.
\label{phyr1}
\end{equation}
Now, if we consider homogeneous initial density as another initial condition, then like the LTB case here also we can write the following expression of Misner-Sharp mass term,
\begin{equation}
F(t_i,r)=\int_0^r\rho(t_i,\tilde{r})\tilde{r}^2d\tilde{r}\,\,=\frac{\rho_i}{3}r^3= M_0r^3\,\, ,
\label{misner1}
\end{equation} 
where $M_0=\frac{\rho_i}{3}$.
As Misner-Sharp mass term is time-independent, the expression of it in eq.~(\ref{misner1}) remains the same throughout the collapse. The initial expression of $G(t,r)$ can be written as $G_i(r)$ which can be chosen.
Let us define another variable $k(t,r)=\frac{P_{\perp}}{\rho}$. Using eqs.~(\ref{misner1}), (\ref{scaling1}), and (\ref{eineqns2})  we get,
\begin{eqnarray}
\label{eq:EEsrhokappaGt}
\rho &=& \frac{3 M_0}{f^2(f + r f^{\prime})}.\label{energy1} \\
k &=& \frac{f}{2}\frac{r \nu^{\prime}}{f + r f^{\prime}}\, ,\label{kvalue1} \\
\dot{G} &=& 2\frac{r \nu^{\prime}}{f + r f^{\prime}} \dot{f}G\,\, ,\label{gvalue}\\
M=M_0 &=& f\left(\frac{1 - G}{r^2} + e^{-2\nu} \dot{f}^2\right).
\label{mvalue}
\end{eqnarray}
As we know, in this collapsing system we always have two free functions to choose. The above equations also indicate the same thing: four equations and six unknowns ($\rho, G, M, k,\nu, f $). Here, we consider $M$ and $k$ as two free functions. Due to the initial homogeneous density profile, we get the expression of the first free function $M$ as $M_0$. However, for $k$ it is very difficult to choose any expression, as we have to maintain all the collapsing regularity conditions which were discussed in previous section. 

As $f(t,r)$ decreases monotonically with respect to $t$, we can write any dynamic variable as a function of $(f,r)$ instead of $(t,r)$, as we did for  dust collapse. Now, the solution of above equations can be written in terms of $(f,r)$ as,
\begin{eqnarray} 
\nu(t,r) &=& 2\int_0^r d\tilde{r}~k\frac{f + \tilde{r} f^{\prime}}{\tilde{r} f} = 2\int_0^r d\tilde{r}~\frac{k}{\tilde{r}} + 2\int_0^r d\tilde{r}~k \frac{f^{\prime}}{f}, \label{nueq1} 
\\
\int \frac{dG}{G} &=& 4\int_f^1  d\tilde{f}~\frac{k}{\tilde{f}} \Rightarrow G(f,r) = G_0(r)~exp{\left[4\int_f^1 d\tilde{f}~\frac{k}{\tilde{f}}\right]}, \label{Geq1} \\
\dot{f} &=& -e^{\nu}\left[\frac{G-1}{r^2} + \frac{M_0}{f}\right]^{1/2} \Rightarrow \frac{\dot{f}}{f}= -e^{\nu}\left[\frac{G-1}{r^2f^2} + \frac{M_0}{f^3}\right]^{1/2}\,\, ,
\label{eq:GCFunctions}  
\end{eqnarray}
where we use $\dot{G}=G_{,f}\dot{f}$ and $G_0(r)$ is integration constant. Using the above expression of $G$ and $\nu$, we can write the general collapsing metric as,
\begin{equation}
ds^2 = -e^{ 2\int_0^r d\tilde{r}~k\frac{f + \tilde{r} f^{\prime}}{\tilde{r} f}}~dt^2 + \frac{(f + r f^{\prime})^2}{G_0(r) e^{4\int_f^1 d\tilde{f}~\frac{k}{\tilde{f}}}}~dr^2 + r^2 f^2~d\Omega^2,
\end{equation}
Now, as we have one degree of freedom left, we can choose a constant value of $k$. For the constant value of $k$, we get the following expression of $\nu$,
\begin{equation}
\label{nu}
\nu(f,r) = 2k_c\ln{r} + 2k_c\int dr~\frac{f^{\prime}}{f}\,\, , \nonumber
\end{equation}
where $k_c$ is the constant value of $k$. So, for any value of $f$, we cannot remove the logarithmic divergence part (the first part of $\nu$ expression ). Therefore, regularity condition will be violated if one considers a constant value of $k$ throughout the collapse.

Now we discuss the equilibrium conditions for the collapsing cloud
and we show how the pressure in the collapsing system can equilibrate the system in an asymptotic time. In the previous section, we have seen that any spherically symmetric dust collapse terminates necessarily into a space-time singularity, where the nature of the singularity, namely visible or otherwise, depends on the initial conditions. However, in the present scenario, due to the non-zero value of pressure, we have an extra degree of freedom which can be used to balance the gravitational pull to achieve an asymptotic equilibrium configuration as an end state of gravitational collapse, as we show below.

In this type of collapsing system we always can define a potential $V(f,r)$ as,
\begin{equation}
V(f,r) = - \dot{f}^2 = - e^{2\int_0^r d\tilde{r}~k\frac{f + \tilde{r} f^{\prime}}{\tilde{r} f}}\left[\frac{G_0 e^{4\int_f^1 d\tilde{f}~\frac{k}{\tilde{f}}}-1}{r^2} + \frac{M_0}{f}\right].
\label{pot}
\end{equation}
When the collapsing system reaches the equilibrium state, we should have $f=f_e$, $\dot{f}=\ddot{f}=0$, where the subscript `$e$` specifies the equilibrium value of any
quantity as,
$$f_e(r) \equiv \lim_{t \to \cal{T}} f(r,t),$$ 
where $\cal{T}$ is very large comoving time. Using this equilibrium condition and eq.~(\ref{pot}), we get,
\begin{eqnarray}
V(f_e, r) &=& 0 \Rightarrow \left[\frac{G_e-1}{r^2} + \frac{M_0}{f_e}\right] = 0, \nonumber \\
V_{,f}(f_e, r) &=& 0 \Rightarrow \left[\frac{(G_e)_{,f}}{r^2} - \frac{M_0}{f_e^2}\right] = 0\,\, ,
\end{eqnarray}
which implies,
\begin{eqnarray} \label{eq:Ge}
G(f_e, r) &=& 1 - \frac{M _0 r^2}{f_e}, \\
G_{,f}(f_e, r) &=& \frac{M_0 r^2}{f_e^2}\,. \nonumber
\end{eqnarray}
Using the above equation and eq.~(\ref{Geq1}), we get,
\begin{equation}
G_e = G_0(r) = 1 - \frac{M_0 r^2}{f_e}\,\,\label{Ge1} .
\end{equation}
Using eq.~(\ref{energy1}), the expression of energy density $\rho_e$ at equilibrium can be written as,
\begin{eqnarray} \label{eq:rhoe}
\rho_e = \frac{3M_0}{f_e^2[f_e + r f^{\prime}(f_e,r)]}\,\, .
\end{eqnarray}
Now, from eq.~(\ref{gvalue}) and eq.~(\ref{eineqns2}) we can write,
\begin{eqnarray}
P_{(\perp)e}=\frac12\rho_e\frac{r\nu^{\prime}_e}{f_e+rf^{\prime}_e}f_e~~ =~~ \frac{\rho_e}{4}f_e\frac{(G_{,f})_e}{G_e}\,\, . 
\label{angpress}
\end{eqnarray}
Using eqs.~(\ref{Ge1}), (\ref{eq:rhoe}), and (\ref{angpress}) we get,
\begin{eqnarray}
k_e = \frac{P_{(\perp)e}}{\rho_e} = \frac{1}{4}\frac{M_0r^2}{f_e - M_0r^2}\,\, .
\label{eq:kappae}
\end{eqnarray}
We can now demand a constant value of $k_e$. As we discussed previously, due to the logarithmic divergence, we cannot choose a constant value of $k$ throughout the collapse. We always can think that this constant $k$ value can be achieved in asymptotic time. As the collapsing system cannot reach this condition in a finite time, the regularity condition will always be maintained. So, here we always consider that the collapsing system will reach the equilibrium condition in a very large comoving time. 

From eq.~(\ref{eq:kappae}), we can see that for $f_e=br^2$ (where $b$ is a positive constant) the value of $k_e$ becomes constant, and it can be written as,
\begin{equation}
k_e=\frac14\frac{M_0}{b - M_0}\,\, .
\end{equation}
For $f_e=br^2$, $G_e$ becomes,
\begin{equation}
G_e=1-\frac{M_0}{b}\,\, .
\label{Gfinal}
\end{equation}
\begin{figure}[t]
\includegraphics[scale=0.5]{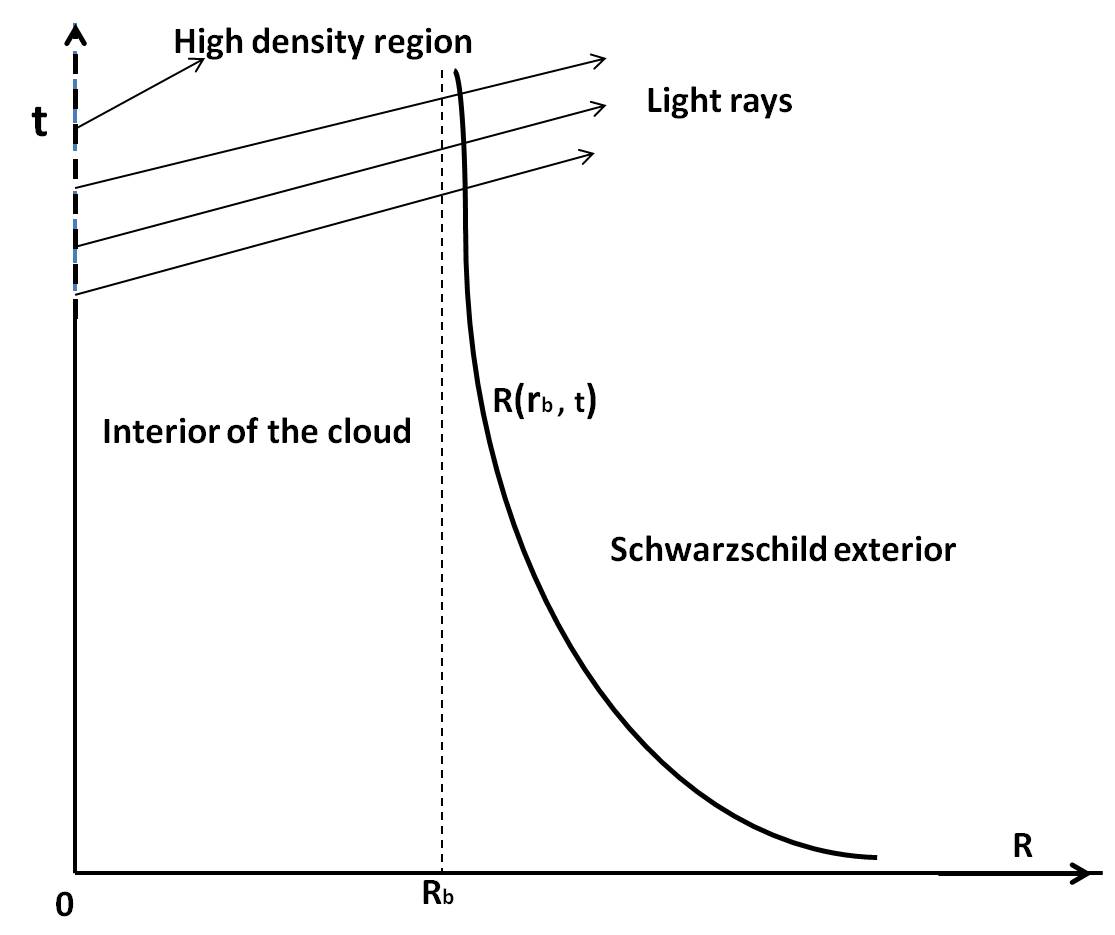}
\caption{The equilibrium configuration is achieved asymptotically
and settles to a final radius. The central density becomes
very large and tends to singularity in infinite co-moving time.}
\label{equilibrium}
\end{figure}
Using eq.~(\ref{nueq1}), we get the final expression of $\nu$ as,
\begin{eqnarray}
\nu_e(r_b) - \nu_e(r) &=& 2k_e \int_r^{r_b} d\tilde{r}~\frac{f_e + \tilde{r} f^{\prime}(f_e,\tilde{r})}{\tilde{r} f_e} \nonumber \\
&=& 2k_e \int_r^{r_b} d\tilde{r}~\frac{(\tilde{r}^2 + 2 \tilde{r}^2)}{ \tilde{r}^3}\nonumber\\
&=& 6k_e \ln{\frac{r_b}{r}} = 2 k_e \ln{\left(\frac{R_e(r_b)}{R_e(r)}\right)}\,\, ,
\label{nub}
\end{eqnarray} 
where $r_b$ is the maximum radius of the final static metric. Here, we use $R_e(r)=rf_e=br^3$. From above eq.~(\ref{nub}), we can write the $g_{tt}(R_e)$ term of final static metric as,
\begin{eqnarray}
g_{tt}(R_e)=e^{2\nu(r_b)}\left(\frac{R_e(r)}{R_e(r_b)}\right)^{4k_e}\,\, .
\label{gtte}
\end{eqnarray}
From eq.~(\ref{Gfinal}) and eq.~(\ref{gtte}), we can now write the final static metric which can be formed at the end state of gravitational collapse in asymptotic time:
\begin{equation}
ds^2=-e^{2\nu(r_b)}\left(\frac{R_e(r)}{R_e(r_b)}\right)^{\frac{M_0/b}{1-\frac{M_0}{b}}}dt^2+\frac{dR_e^2}{\left(1-\frac{M_0}{b}\right)}+ R_e^2~d\Omega^2\,\, .
\end{equation}
As for the present case we do not consider any matter flow across the boundary of collapsing matter cloud, the final space-time should match with an external Schwarzschild metric:
\begin{equation}
ds^2=-\left(1-\frac{\mathbb{M}_0 \mathbb{R}_b}{\mathbb{R}}\right)dt^2+\frac{d\mathbb{R}^2}{\left(1-\frac{\mathbb{M}_0 \mathbb{R}_b}{\mathbb{R}}\right)}+\mathbb{R}^2d\Omega^2\,\, ,
\end{equation}
where $\mathbb{R}(r_b)=\mathbb{R}_b$. From induced metric matching and extrinsic curvature matching, we get the following conditions,
\begin{eqnarray}
\mathbb{R}(r_b)&=&R_e(r_b)\,\,\,\,\,\,\Rightarrow\,\,\,\,\, \mathbb{R}_b=R_b\,\,\, ,\\
G_{e}(r_b)&=&1-\frac{M_0}{b}=1-\mathbb{M}_0\,\,\, ,\\
e^{2\nu(r_b)}&=&1-\mathbb{M}_0\,\,\ .
\end{eqnarray}
From the matching conditions, one can see that the Schwarzschild mass $\mathbb{M}_0=\frac{M_0}{b}$. Satisfying all the above matching conditions, we get the expression of final metric: 
\begin{eqnarray}
ds^2&=& -\left(1-\frac{M_0}{b}\right)\left(\frac{R_e}{R_b}\right)^{\frac{M_0/b}{1-\frac{M_0}{b}}}~dt^2 + \frac{dR_e^2}{\left(1-\frac{M_0}{b}\right)}+ R_e^2~d\Omega^2\,\, .
\end{eqnarray}

This space-time is called the JMN space-time
\cite{Joshi:2011zm}, which has a singularity at the center 
but no trapped surfaces, and which is obtained as a limiting spacetime from gravitational collapse. The collapsing space-time can equilibrate itself to this static space-time in an asymptotic time. Now, the ratio between the Misner-Sharp mass and physical radius becomes,
\begin{equation}
\frac{F(r)}{R_e}=\frac{M_0}{b}\, .
\end{equation}
As it was shown previously, for a dynamic space-time the ratio between Misner-Sharp mass and physical radius should be always less than unity to avoid apparent horizon or trapped surface formation. From the $g_{rr}$ component of JMN space-time, it can be checked that $\frac{M_0}{b}$ should always be less than one. Therefore, there is no trapped surface or apparent horizon. Light rays from the central high-density region can reach the asymptotic observer. In fig.~(\ref{equilibrium}) we diagrammatically describe the global visibility of ultra-high density region at the center.

From the above discussion, one can see that a collapsing system can have a stable configuration as a final state if there exists some pressure in it.  The internal pressure can balance the gravitational pull to slow down the collapse and generate the possibility of an equilibrium state with a very high-density region at the center. Besides this equilibrium state, a black hole can also be the possible final state of non-dust-like collapse. The final outcome depends upon the initial conditions of the collapsing scenario. The collapsing process, which we have discussed above, is a `slow process' in the sense that it takes a very large time to equilibrate the collapsing system. On the other hand, the collapse processes, which terminate into black-holes or naked singularity, generally take a finite amount of time to reach the final state.
The main reason behind this may be that when trapped surfaces form at any radius of collapsing dust-like matter, they propagate all over the collapsing system or may suck up all the matter into it in a finite amount of time. However, when there is no trapped surface throughout the collapse, the presence of pressure in the matter cloud slowly balance the gravitational pull and it takes a very large amount of comoving time to equilibrate the collapsing system. If we recall, in the virialization process also, the collapsing system with gravitationally interacting $N$ number of particles can be virialized in only a very large system time.

Therefore, on the basis of both these processes, namely the general relativistic process of equilibrium, and the virialization process, we find that the equilibrium configuration is achieved only in a very large limit of the  comoving time general relativistically, or the system time classically. We may therefore conclude or indicate that the above-mentioned relativistic process of equilibrium is a general relativistic analog of usual virialization. As baryonic matter can dissipate its energy, the structures of baryonic matter are much more dynamic compared to the structure of dark matter \cite{white}. Local dissipations can change the stable configuration of baryonic matter in a finite time and on the other hand, dissipation-less dark matter retains its stable structure for a very large time. Therefore, the above-mentioned general relativistic technique of equilibrium is suitable for describing dark matter collapse.
In the next section, we will use this general relativistic technique to describe how dark matter forms its halo structures.
\section{Primordial Dark-matter Halo Formation}
\label{Halo}
As we know, dark matter created the first structures in the universe. As the dark matter decoupled early, the density perturbations in dark matter field started to grow during the time when the baryonic matter was in thermal equilibrium. Gradually, the density perturbations
became such that the density contrast $\Delta \rho/\bar{\rho} \sim 1$, and the dynamics of perturbation modes entered into the non-linear regime. As a consequence of the nonlinear growth of dark matter density contrast,  various patches of overdense regions were formed. For simplicity, those overdense patches can be assumed as spherically symmetric overdense matter distributions. As the background, namely the flat Friedmann-Lemaitre-Robertson-Walker (FLRW) universe expands in a homogeneous and isotropic fashion, these overdense regions also expand initially in an isotropic and homogeneous fashion. Gradually, each of these overdense patches detaches from the cosmic expansion and behave like a sub-universe. At a certain time, these patches stop expanding and then start to collapse under their own gravity. The total dynamics of the overdense regions can be modeled by the dynamics of a homogeneous, spherically symmetric dust cloud. Conventionally, closed FLRW metric is used to model the evolution of these overdense regions. This model is known as spherical top hat collapse \cite{Gunn}. We know that any homogeneous dust collapse finally terminates into a black hole. Therefore, in this model virialization technique is used to stabilize the collapsing matter.

\subsection{Top-Hat Collapse Model}
In the spherical top-hat collapse
model, the metric of the overdense regions is given by the closed FLRW space-time \cite{Fried}:
\begin{eqnarray}
 ds^2 = - dt^2 + {a^2(t)\over 1- r^2 }dr^2 + a^2(t) r^2 d\Omega^2\,,
\label{FLRWMetric}
\end{eqnarray}
where $a(t)$ is the scale-factor of the closed FLRW space-time and the range of radial coordinate is $0\le r \le 1$. As in top-hat collapse model the matter inside the overdense regions are considered as dust, the above metric can be thought as a particular form of LTB metric, where the physical radius $R(t,r)= r a(t)$, the Misner-Sharp mass $F(r)= F_0 r^3$ and $E(r)=-r^2$. Here $F_0$ is related with initial density of overdense region.  From the expression of Misner-Sharp mass in eq.~(\ref{Misner}), we can write,   
\begin{equation}
\frac{F_0}{a(t)}=1+\dot{a}^2
\label{topmisner}
\end{equation}
The above equation of motion can be written in a more compact form as,
\begin{eqnarray} 
\frac{H^2}{H_0^2}=\Omega_{m0}\left(\frac{a_0}{a}\right)^3 + (1-\Omega_{m0})\left(\frac{a_0}{a}\right)^2 \, ,
\label{fried}
\end{eqnarray}
The above differential equation is known as the Friedmann equation,
where $H=\dot{a}/a$ is the Hubble parameter for the overdense
sub-universe and $H_0,\,a_0$ are  the initial values of $H$ and $a$. Here, the initial values correspond to the time when the overdense regions start to evolve independently and detach from the background cosmic expansion. Here, $\Omega_{m0}=\rho_0/\rho_{c0}$, where $\rho_{c0} =
3H_0^2$, and $\rho_0$ is the initial homogeneous matter density. So from eq.~(\ref{F1}), we can write: $F_0=\frac{\rho_0}{3}$.
The solution of eq.~(\ref{fried}) can be written in a parametric form :
\begin{eqnarray}
 a = \frac{a_m}{2}(1- \cos\theta),~~~
 t = \frac{t_m}{\pi}(\theta-\sin\theta)\, ,
\label{at}
\end{eqnarray}
where $t_m$ is the time when the scale factor $a(t)$ reaches its maximum limit $a_m$. When $\theta=\pi$, we get the maximum value of $a(t)$. We can write $a_m$,$t_m$ in terms of $\Omega_{m0}$, $H_0$ and $a_0$ as,
$$a_{m}={a_0 \Omega_{m0}\over (\Omega_{m0}-1)}\,,\,\,\,\, t_{m}={\pi \Omega_{m0}\over 2 H_0 (\Omega_{m0}-1)^{3/2}}\,,$$
where we always have $\Omega_{m0}>1$ for the overdense region. In top-hat collapse model, the overdense regions are considered as sub-universes. These sub-universes, which are described by closed FLRW metric, expand first and reach a maximum scale factor limit at time $t_m$. At this point the spherical overdense regions start collapsing. One can calculate the ratio between the the density of overdense region and the density of background at the turn-around point $t=t_m$:
\begin{eqnarray}
\frac{\rho(t_{m})}{\bar{\rho}(t_{m})} = \frac{9\pi^2}{16} \sim 5.55\,, 
\label{impfac}
\end{eqnarray}
which implies that the spherically symmetric overdense region starts collapsing when its density becomes $5.55$ times higher than the density of the background. As we know from the LTB collapse analysis, any spherical dust collapse inevitably terminates into a space-time singularity. For the present case, the ratio between Misner-Sharp mass and physical radius is:
\begin{eqnarray}
\frac{F(r)}{R(r,t)}=r^2\frac{F_0}{a(t)}=r^2(1+\dot{a}^2)\,\, ,
\label{apparent1}
\end{eqnarray}
where we use eq.~(\ref{topmisner}). From the above equation one can check that trapped surfaces form in this type of collapse and as time goes by, they gradually propagate from the edge ($r=1$) of the sub-universe to the center ($r=0$). So, the final singularity at $\theta= 2\pi$ must be covered by an event horizon, and therefore a Black hole should be the final state for this type of collapse. To avoid the Black hole at the end of this kind of
gravitational collapse, the virialization technique is used to equilibrate the collapsing system. As we know, a gravitating system virializes when the following condition is full-filled,
\begin{equation}
\langle T \rangle=-\frac12 \langle V_T \rangle \, ,
\end{equation}
where $ \langle T \rangle$ and $\langle V_T \rangle$ are the average kinetic and potential energy respectively.

Now, in spherical top-hat collapse model when the spherically symmetric overdense region having total mass $M$ reaches its maximum scale factor limit, it momentarily stands still before starting to collapse. At that moment kinetic energy is zero, and all the energy is potential energy,
\begin{equation}
V_T=-\frac{3M^2}{5R_m}\, ,
\end{equation}
where $R_m$ is the maximum physical radius of the overdense region at time $t_m$. When it has collapsed to $R=\frac{R_m}{2}$, then using the conservation of total energy one can derive the expressions of potential energy and kinetic energy,
\begin{eqnarray}
V_T&=& -\frac{6M^2}{5R_m}\, ,\nonumber\\
T&=& \frac{3M^2}{5R_m} = -\frac{V_T}{2}\, .
\end{eqnarray}
At this point, it is considered that the collapsing system virializes itself to a static configuration.
So, in top-hat collapse model the overdense regions virialize when the scale factor becomes
$a_{\rm vir}=\frac12 a_{m}$. From  eq.~(\ref{at}), one can show that this virialization will happen when $\theta = \frac{3\pi}{2}$.  At this stage the ratio is, 
\begin{eqnarray}
  \frac{\rho(t_{\rm vir})}{\bar{\rho}(t_{\rm vir})} \sim 145\,.
\label{ratio2}
\end{eqnarray}
So, the top-hat collapse model tells us that at the time $t_{\rm vir}$, when  the dark matter virializes and forms its stable halo structure, the density of the halo becomes $145$ times larger than the background. Though the top-hat collapse is a very simple model to describe the non-linear evolution of density perturbation modes, it gives numbers which are very important for astrophysics.

We note, however, that there are many problems with this model as it is a totally homogeneous, spherical dust collapse. The collapsing system, which is described by the top-hat model, starts collapsing with a homogeneous matter field and remains homogeneous throughout the collapse, which is unrealistic. The second issue with this model is that here a general relativistic technique is used to describe the dynamics of the overdense regions, however, to describe a stable configuration of the collapsing system a Newtonian virialization technique is used.  Therefore it follows that these two techniques are not glued together properly, neither they are compatible with each other. As we know, virialization happens in gravitating systems through different processes for which it takes a large amount of system time. It just cannot happen in a short finite amount of time or a moment of time. Before $t=t_{\rm vir}$ there is no sign of stabilizing process. Before that time the system undergoes a catastrophic gravitational collapse which could lead to a space-time singularity, but suddenly for the sake of stabilization, the virialization process is used. There is no doubt that the collapsing system  would virialize when the system radius $R\sim\frac{R_m}{2}$. However, the technique, which is used in the top-hat collapse model, cannot describe the equilibrium process properly. 

Our purpose here is to point out that, one can generalize the top-hat collapse model,
retaining some of its simplicities and attractive features. In the next section, we are going to discuss  a modification of the top-hat collapse model where we do not need to introduce a Newtonian virialization technique during the collapse to describe a stable final state, but we can use a general relativity model.  
In this sense, we can say that this offers a general relativistic alternative to the Newtonian virialization technique. While the model presented here is still very much a toy model which may not be physically realistic, we believe it provides several useful insights towards achieving a general relativistic description for structure formation in the universe. We also indicate here a few possibilities towards the further generalization of this scenario presented.

\subsection{Modification of Top-Hat collapse model}
In this section, we will briefly review the literature \cite{Bhatt}, where a special technique is used to modify the top-hat collapse model. As we know, in the top-hat collapse model the collapsing matter field is homogeneous and pressureless throughout the whole collapse process, and therefore we cannot achieve any equilibrium configuration general relativistically. Therefore, in that model, an ad-hoc input of virialization is introduced and needed to stabilize the collapsing system. On the other hand, in the previous section, we have discussed a collapsing scenario which leads to a stable configuration at the end stage of gravitational collapse. The dynamics of that scenario is governed by the general collapsing metric,  
\begin{eqnarray}
ds^2 = - e^{2\nu(r,t)} dt^2 + {R'^2\over G(r,t)}dr^2 + R^2(r,t) d\Omega^2\, .
\label{genmetric2}
\end{eqnarray}
With this metric, one can describe a matter field which has non-zero pressure in it. The pressure term inside the matter field is responsible for the final equilibrium configuration. However, it is very difficult to describe the total evolution of spherical overdense regions using the general collapsing metric, as we need the initial functional expressions for all the functions $\nu(r,t)$, $R(r,t)$ and $G(r,t)$. In the cosmological scenario, when the overdense regions detach from background expansion, it becomes a non-trivial task to set the initial form of those functions.

One can simplify the situation by considering the general collapsing metric in the collapsing phase only. This will help us to find the initial conditions. In this technique, the initial expansion phase of homogeneous, pressureless fluid remains as it was in the top-hat collapse model. However, from the turn-around point, the collapsing phase is described by  the general collapsing metric. During the expansion phase, the matter field is homogeneous and pressureless carrying some properties of background expansion which is described by a flat FLRW metric. Now, from the starting point of the collapsing phase, inhomogeneity and pressure start to grow gradually. This pressure term finally equilibrates the collapsing system in the same way as it is described in the previous section. To pursue this total evolution of the overdense matter field, we need to glue closed FLRW metric with general collapsing metric at the turn-around point. This can be done by using Darmois-Israel junction conditions \cite{poiss}, \cite{Israel:1966rt}, \cite{Darmois1927} on the spacelike hypersurface, namely, $\Phi(t)=t-t_{m}=0$. Therefore, from the junction conditions we can get the initial functional expressions of the unknown functions in the general collapsing metric, and then we can use the general relativistic equilibrium technique to investigate the final outcome. So, in the new model of spherical collapse, the overdense sub-universe initially expands in an isotropic and homogeneous fashion (with zero pressure), which is described by closed FLRW metric. Then from the turn-around point, the non-zero pressure and inhomogeneity develop gradually in the matter field, and general collapsing metric is then used to describe this later phase.

In the top-hat collapse model, as the  collapsing matter is pressureless, the mathematical technique for describing the dynamics of the overdense sub-universe is very simple. However, as the collapsing matter field is homogeneous and pressureless throughout the collapse, the model becomes physically unrealistic and an ad-hoc situation and virialization technique is used to stabilize the collapsing system. In the modified version of the top-hat model, the matter is more realistic and a new general relativistic approach is used to describe the total evolution of overdense sub-universes.

\subsection{Inhomogeneous and Non-dust like Spherical Collapse}
\label{nondust}
Here we discuss the modified top-hat model in some detail. To achieve a stable configuration general relativistically, we will glue the closed FLRW metric with general collapsing metric at the turn-around point by using Darmois- Israel junction conditions. These conditions will give us some initial conditions which will be used to investigate the final fate of spherical gravitational collapse of inhomogeneous, non-dustlike fluid.
\begin{figure}[t]
\begin{center}
\includegraphics[width=2.5in]{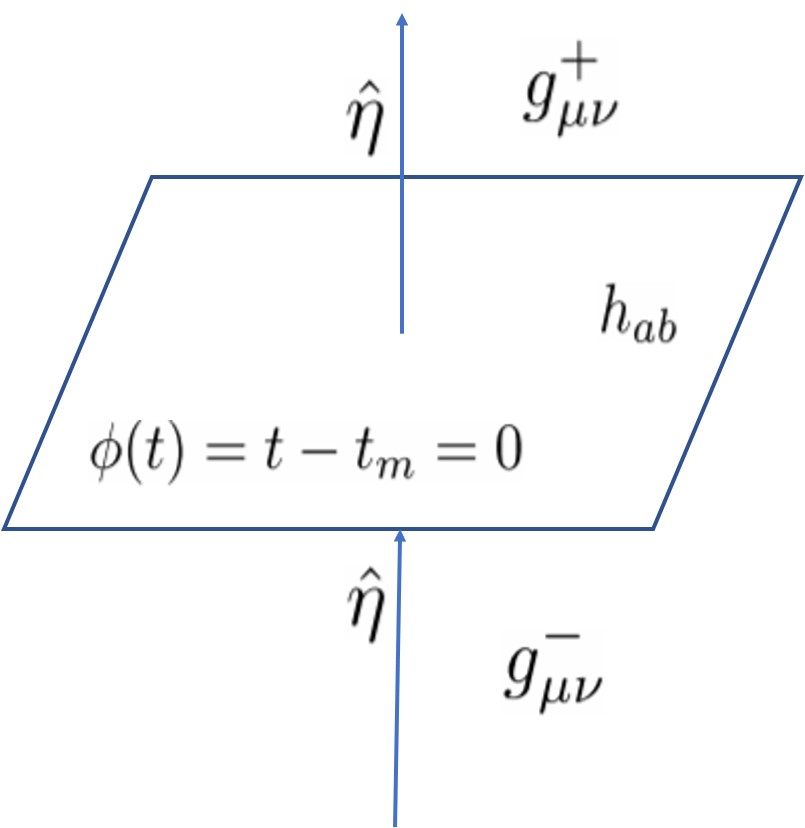}
\caption {Here $g_{\mu\nu}^+$ is contracting space-time and $g_{\mu\nu}^-$ is the closed FLRW metric.}
\label{hypersurfaces1}
\end{center}
\end{figure}

\subsubsection{Matching conditions}
We have to match closed FLRW metric with a general collapsing metric at the spacelike hypersurface: $\Phi(t)=t-t_m=0$. In this case, for smooth matching, according to the junction conditions, we need to match induced metrics and extrinsic curvatures at the constant time spacelike hypersurface. Here the two induced metrics will be denoted as $g_{\mu\nu}^+$ and $g_{\mu\nu}^-$, where `+' and `-' are related with general collapsing metric and closed FLRW metric respectively. In fig.~(\ref{hypersurfaces1}) we schematically describe the space-time structure. Now, using eq.~(\ref{genmetric2}) and eq.~(\ref{FLRWMetric}), we can write the induced metrics as,
\begin{eqnarray}
ds^2_{+}=\left(\frac{R^{\prime 2}(r,t_m)dr^2}{G(r,t_m)}+R^2(r,t_m)d\Omega^2\right),~~
ds^2_{-}=\frac{a_m^2dr^2}{1-r^2}+r^2 a_m^2d\Omega^2~.
\label{nondustmatching}
\end{eqnarray}
From the first condition of smooth matching, we have 
\begin{eqnarray}
g_{\mu\nu}^{+}(t\rightarrow t_{m})&=&g_{\mu\nu}^{-}(t\rightarrow t_{m})\,\, ,
\end{eqnarray}
and this will give us the initial expressions of $G(r,t)$ and $R(r,t)$ as,
\begin{eqnarray}
R(r,t_{m})=rf(r,t_{m}) = ra_{m},~~ G(r,t_{m}) = 1-r^2\, .
\label{nmatch}
\end{eqnarray}
The first initial condition implies, $f(r,t_m)=a_m$. The second condition for smooth matching is
\begin{eqnarray}
K^{+}_{\mu\nu}(t\rightarrow t_{m})&=&K^{-}_{\mu\nu}(t\rightarrow t_{m})\,\, ,
\end{eqnarray}
where $K_{\mu\nu}$ is the extrinsic curvature, which can be mathematically written as,
$ K_{ab} = {\eta_{\alpha;}}_\beta e^\alpha_a e^\beta_b\,\, , $ where semicolon indicates co-variant derivative.
Here, $\eta$ is a vector perpendicular to the $\Phi(x_\alpha)=0$ spacelike hypersurface, and it is defined as 
\begin{eqnarray}
 \eta_\mu = {\epsilon\Phi,_\mu\over |g^{\alpha\beta}\Phi,_\alpha\Phi,_\beta|^{1/2}}\, 
\label{norm}
\end{eqnarray}
where $e^\alpha_a$ are the  tangents to that spacelike hypersurface. Comma in the above expression indicates ordinary partial derivative, and $\epsilon\equiv +1$ is used for timelike hypersurface and $\epsilon\equiv -1$ is for spacelike hypersurface. Here, $e^\alpha_a$ is defined as, $e^\alpha_a\equiv\frac{\partial x^{\alpha}}{\partial l^a}$, where $l^a$ is the induced coordinate system on the hypersurface.   As mentioned, in
our case the hypersurface is $\Phi(t)=t-t_{m} =0$. On this spacelike hypersurface the induced coordinate system is, $l^a=\left\lbrace r,\theta,\phi\right\rbrace$. So, the components of tangent vectors from both sides can be written as,
\begin{equation}
e_r^{\alpha\pm}=\left\lbrace 0,1,0,0\right\rbrace\,\, ,\,\,e_{\theta}^{\alpha\pm}=\left\lbrace 0,0,1,0\right\rbrace\,\, ,\,\,e_{\phi}^{\alpha\pm}=\left\lbrace 0,0,0,1\right\rbrace\,\, .
\end{equation}
The normal vectors to the spacelike hypersurface from the both sides can be written as,
\begin{eqnarray}
\eta^+_{\mu}=\left\lbrace -e^{\nu(r,t)},0,0,0\right\rbrace\,\,\, ,\,\,\eta^-_{\mu}=\left\lbrace -1,0,0,0\right\rbrace\,\, .
\end{eqnarray}
Using the above expressions of tangents and normals to the spacelike hypersurface $\Phi(t)$, we can now write the expressions of non-zero extrinsic curvature components for the both sides. For closed FLRW metric we get,
\begin{eqnarray}
K_{rr}^{-}&=&\frac{a(t)\dot{a}(t)}{1-r^2}\,\, ,\nonumber\\
K_{\theta\theta}^{-}&=&\frac{K_{\phi\phi}^{-}}{\sin^2\theta}=r^2a(t)\dot{a}(t)\,\, .
\label{extrin1}
\end{eqnarray}
As the scale factor $a(t)$ of closed FLRW metric reaches its maximum limit $a_{m}$ at $t = t_{m}$,  $\dot{a}(t_{m})=0$. Therefore, on the spacelike hypersurface all the extrinsic curvature components for closed FLRW metric become zero. So we can write,
\begin{eqnarray}
K_{rr}^{-}|_{t=t_m}=K_{\theta\theta}^{-}|_{t=t_m}=K_{\phi\phi}^{-}|_{t=t_m}=0.
\label{extrin2}
\end{eqnarray}
Now, for the general collapsing metric we get the following non-zero components of extrinsic curvature,
\begin{eqnarray}
K_{rr}^+ &=&\frac{e^{-\nu (r,t)}}{2 G(r,t)^2}\left(r f'(r,t)+f(r,t)\right)\left\lbrace-\left(r
   f'(r,t)+f(r,t)\right) {\dot G}(r,t)+2 \left({\dot f}(r,t)+r{\dot f}'(r,t)\right) G(r,t)\right\rbrace\,\, ,\nonumber\\
K_{\theta\theta}^+ &=& +r^4 f(r,t) {\dot f(r,t)} \left(e^{-\nu (r,t)}\right) = \frac{K_{\phi\phi}^+}{\sin^2\theta}\,\, .
\end{eqnarray}
Now, from the induced metric matching we get $f(r,t_m)=a_m$, which indicates that $f(r,t)$ becomes $r$ independent at time $t=t_m$. From the matching of  azimuthal component of extrinsic curvature we can get the condition: $\dot{f}(r,t_m)=0$. Using these two conditions, and from the matching of radial component of extrinsic curvature, we get $\dot{G}(r,t_m)=0$. So, from the matching conditions we get the following initial conditions for collapse,
\begin{eqnarray}
f(r,t_m)=a_m\,\, ,\,\,\, G(r,t_{m}) = 1-r^2\,\, ,\,\,\, \dot{f}(r,t_m)=0\,\, ,\,\,\,\dot{G}(r,t_m)=0\,\, .
\label{initial1}
\end{eqnarray}

%\caption {$g_{\mu\nu}^+$ is contracting space-time and $g_{\mu\nu}^-$ is the closed FLRW metric, and $t_{\rm max}$ is denoted by $t_m$.}
%\label{hypersurfaces1}
%\end{center}
%\end{figure}

\subsubsection{Inhomogeneous Anisotropic Collapse}
Now, using the above initial conditions, we can investigate the final equilibrium metric, using the technique which we have described in the previous section. Here, the collapsing matter is considered as inhomogeneous, non-dustlike. For a simple case, one can  consider dustlike inhomogeneous fluid during collapse after the expansion of the homogeneous, dustlike fluid. For this simple case we need, to match closed FLRW metric with LTB metric on the spacelike hypersurface $\Phi=t-t_m=0$. However, it can be shown \cite{Bhatt} that if the fluid remains dustlike during the expansion as well as the collapsing phase, and if the fluid is homogeneous during the expansion phase, then the fluid has to be homogeneous during the collapsing phase also. Therefore, inhomogeneous dust collapse after homogeneous dust expansion is not possible. Now, we are going to discuss the collapsing scenario where we consider the collapse of an inhomogeneous, anisotropic fluid after the expansion of the homogeneous dustlike fluid.
%As we have argued, the general relativistic technique of equilibrium can be thought as a general relativistic analog of virialization process. Therefore, 

For the anisotropic collapse, we will first consider one of the simplest examples where the fluid has zero radial pressure throughout the collapse. As we know, zero radial pressure gives us the time-independent Misner-Sharp mass term, and that solution can be derived from the initial conditions. Using the Misner-Sharp mass expression in eq.~(\ref{ms}) and the initial conditions in eq.~(\ref{initial1}), we can get following functional form of Misner-Sharp mass,
\begin{eqnarray}
F(r) \equiv F(r,t_{m}) = rf(r,t_{m})\left[1- G(r,t_{m}) +
r^2 e^{-2\nu(r,t_{m})} \dot{f}^2(r,t_{m})\right]= a_{m} r^3\,,
\label{msm}
\end{eqnarray}

As was found previously, the equilibrium condition is achieved when the following condition holds,
\begin{eqnarray}
\dot{f}_e(r) = \ddot{f}_e(r) =0\,.
\label{stabc2}
\end{eqnarray}
This condition is achieved in a large coordinate time $\mathcal{T}$.  As we mentioned previously, in this collapsing system, we have two degrees of freedom to choose two unknown free functions. This is true for the equilibrium state also. Now, in this case, the Misner-Sharp mass term $F$ is time-independent, and therefore the functional expression of $F$ is fixed from the initial condition of collapse. So, we need to choose the expression of one unknown function. In the previous section, we choose the functional expression of $k_e(r)$ which is the ratio between azimuthal pressure and energy density. Here, we choose the functional expression of the physical radius as
\begin{equation}
R_e(r)=br^{\alpha +1}\, .
\label{afer}
\end{equation}
With this consideration, the system becomes totally solvable and one can derive the expressions of other functions. 

In this case,  for $\alpha=2$, the final metric will not be the same as it was in the previous section. In the previous section, we saw that $\alpha=2$ gives us constant value of $k_e(r)$. In the top-hat collapse model, the overdense region is considered as an isolated universe, and the comoving radius ranges from zero to one. This type of restriction on comoving radius was not there in the previous case. Scaling $a_{m}$ to unity, we first note that as it is a collapsing process, $R(r,t)$ should always be decreasing with time. Therefore, we always have $R_{m} > R_e$ which implies that for any value of $r$,
\begin{equation}
\frac{R_{m}}{R_e} = \frac{1}{br^{\alpha}} > 1 \Rightarrow b < 1~.
\label{ns1}
\end{equation}
As the radial coordinate $r$ can have values very close to zero, the second inequality in the above equation is necessary.
Next, if we demand no apparent horizon at the starting point of the collapse, then the following inequality should hold,
\begin{equation}
\frac{F(r,t_{m})}{R_{m}} = \frac{r^3 a_{m}}{r a_{m}} < 1~,
\end{equation}
which is always satisfied in the range of $r$. As we do not know the analytic solution of total gravitational collapse, we cannot say anything about $\frac{F(r)}{R(r,t)}$ during the collapse.   However, it is interesting to investigate whether there is a possibility of trapped surfaces at the equilibrium state of collapse. When the collapsing matter cloud achieves the equilibrium configuration, the ratio between the Misner-Sharp mass and physical radius becomes, 
\begin{equation}
\frac{F(r)}{R_e} = \frac{r^2}{br^{\alpha}}=1 \Rightarrow r = b^{\frac{1}{2-\alpha}}~.
\label{aphor1}
\end{equation}
From the above equation, one can see that $\alpha\geq 2$ must be ruled out to avoid apparent horizon and trapped surfaces. So, we can say that for a physically realistic gravitational collapse, which has no apparent horizon at the final equilibrium state, the following inequality should hold,
\begin{equation}
0\leq\alpha < 2,~~0<b<1~.
\end{equation}
Therefore, we will use these inequalities in the derivation of the final stable metric.
Now, as we know the equilibrium expressions of $F(r)$ and $R(r,t)$, we can get the final equilibrium space-time using the technique which has been discussed in the previous section. The final static metric is,
\begin{equation}
ds_e^2 = -{\mathcal A}\left(b-r^{2-\alpha }\right)^{\frac{\alpha +1}{\alpha -2}}dt^2 + \frac{(\alpha +1)^2 b^3 r^{2 \alpha }}{b- r^{2-\alpha }}dr^2
+ b^2 r^{2\left(\alpha +1\right)}d\Omega^2~.
\label{meteq}
\end{equation}
Now, if we write the above metric in terms of physical radius, we get,
\begin{equation}
ds_e^2 = -\frac{R_b - \left(\frac{R_b}{b}\right)^{\frac{3}{1+\alpha}}}{R_b}
\left(\frac{1 - \frac{1}{b}\left(\frac{R_b}{b}\right)^{\frac{2-\alpha}{1+\alpha}}}{1 -\frac{1}{b} \left(\frac{R_e}{b}\right)^{\frac{2-\alpha}{1+\alpha}}}\right)^{\frac{1+\alpha}{2-\alpha}}dt^2 
+ \frac{dR_e^2}{1-\frac{1}{b}\left(\frac{R_e}{b}\right)^{\frac{2-\alpha}{1+\alpha}}} + R_e^2d\Omega^2~,
\label{dsfinal}
\end{equation}
where the above metric is matched with Schwarzschild metric at the physical radius $R_b$. 
From eq.~(\ref{meteq}), it can be seen that the metric is valid only in the range: $0<r < b^{1/(2-\alpha)}$. Therefore, we need to match this space-time with the Schwarzschild metric in that range. If we set $\alpha=0$, we get the following space-time,
\begin{equation}
ds_e^2 = -\frac{\left(1 - \frac{R_b^2}{b^3}\right)^{\frac{3}{2}}}{\sqrt{1 - \frac{R_e^2}{b^3}}}dt^2 + \frac{dR_e^2}{1-\frac{R_e^2}{b^3}} + R_e^2d\Omega^2~.
\label{Flo}
\end{equation}
The above space-time is known as  Florides space-time \cite{Florides}. It can be seen that there is no
curvature singularity at $R_e = r = 0$. At $R_e = b^{3/2}$ this space-time has a curvature singularity. Therefore, this solution is totally regular in the range $0\leq R_e<b^{3/2}$. The energy density is homogeneous in this case. 
%Importantly, we have shown here thata regular interior solution like the one discovered by Florides can arise out of a collapse process in a cosmological setting. 

In eq.~(\ref{dsfinal}) we present a new class of space-times which can be formed as an end state of gravitational collapse of inhomogeneous, anisotropic fluid. As we have discussed, this type of equilibrium configuration can develop from a gravitational collapse in very large co-moving time. Therefore, when the overdense regions of dark-matter start collapsing under its own gravity, the non-zero pressure  within  can stabilize the collapsing system in a very large comoving time. As we know, the dark matter is considered as a pressureless fluid in cosmological scale. However, in the halo scale, it need not be always pressureless during the collapse. If the general relativistic technique of equilibrium is considered as a relativistic analog of virialization, then one can always use this technique to explain the total evolution of overdense dark matter regions without using an ad-hoc virialization technique. It can also be shown that the presence of isotropic pressure can lead to a homogeneous, regular space-time as a final state of gravitational collapse. In \cite{Bhatt}, a full analytic solution of this type of gravitational collapse is presented.

\section{conclusion}
We have explored here the general relativistic possibility which
can replace the usual virialization technique that is normally used in the
so-called top-hat collapse models in the structure formation scenarios
in astrophysics and cosmology. We show that while making the departures
from the homogeneous dust form of matter, and when we consider the
more realistic collapse with non-zero pressures, it naturally follows that
we could generate equilibrium configurations which may represent
larger structures. 

As we discussed in the previous section, baryonic matter can create stable structures and one can always describe the collapse process by using the above-mentioned general relativistic technique. However, usually, the stable structures of baryonic matter are not stable forever. Baryonic matter can dissipate its energy and can accumulate into the central region of a self-gravitating system. Therefore, the structures of baryonic matter are far more dynamic than dark matter structures \cite{nfwprof},\cite{Navarro:1996gj},\cite{Jenkins:2000bv}, \cite{white}-\cite{Birnboim:2003xa}. Baryonic matter can form compact objects like stars, planets and also can form black holes, naked singularities in a finite amount of time. On the other hand, dark matter cannot dissipate its energy, and therefore it can form its structure at larger scales like halo scale, galaxy cluster scale etc.  Therefore, one can use the general relativistic technique of equilibrium  more efficiently to describe dark matter collapse. A small pressure in the dark-matter field can create stable structures in asymptotic time as we showed. 

Our model presented here is very much of a toy model in that we
have, for the sake of simplicity, chosen the radial pressures to be
vanishing, which will be typically non-zero for any physically realistic gravitational systems. We note that collapse scenarios that result into equilibrium configurations, and which have non-zero radial pressures, have been developed
in \cite{Joshi:2013dva}. These may be explored towards astrophysically more realistic models \cite{Sayan}. Also, we have not examined here the comparison of actual astrophysics observations with our model, and in terms of numbers. We plan to take up these issues in future work. The advantage of our model, however, is that we have presented here a fully consistent general relativistic model to develop
equilibrium structures such as galaxies, rather than using a patchwork of Newtonian and relativity techniques.

\end{document}